\def\dfrac#1#2{{\displaystyle{#1\over#2}}}

\def\d{\mbox{d}}
\documentstyle[pre,aps,amssymb,epsf,floats]{revtex}
\begin{document}
\draft

\twocolumn[\hsize\textwidth\columnwidth\hsize\csname
@twocolumnfalse\endcsname

\title{Dilute Bose gas: short-range particle correlations and
ultraviolet divergence}

\author{A. Yu. Cherny$^{*}$ and A. A. Shanenko$^{**}$}

\address{Bogoliubov Laboratory of Theoretical Physics,
Joint Institute for Nuclear Research, 141980, Dubna, Moscow region,
Russia}

\date{November 28, 2000}

\maketitle

\begin{abstract}
The modified Bogoliubov model where the primordial interaction is replaced by
the $t$ matrix is reinvestigated. It is shown to provide a negative value of
the kinetic energy for a strongly interacting dilute Bose gas, contrary to
the original Bogoliubov model. To clear up the origin of this failure, the
correct values of the kinetic and interaction energies of a dilute Bose gas
are calculated. It is demonstrated that both the problem of the negative
kinetic energy and the ultraviolet divergence, dating back to the well-known
paper of Lee, Yang and Huang, is connected with an inadequate picture of the
short-range boson correlations. These correlations are reconsidered within
the thermodynamically consistent model proposed earlier by the present
authors. Found results are in absolute agreement with the data of the
Monte-Carlo calculations for the hard-sphere Bose gas.
\end{abstract}
\vskip1pc]
\narrowtext

\section{Introduction}
\label{1sec}
The observations of the Bose-Einstein condensation in the
magnetically trapped alkali atoms~\cite{anddavbrad} significantly
renewed interest in the theory of the Bose-Einstein condensation
(see, e.g., Ref.~\cite{RMP}) and also motivated its extensive
re-examinations. The standard method, which is widely used to study
the Bose gas of neutral atoms, operates with replacing a
pairwise interatomic potential by the effective, or ``dressed", one
(this is why below the name ``effective-interaction" is used for the
approach whose various representations are listed in Ref.~\cite{Lee}).
Such a replacement allows one to overcome calculational
obstacles coming from the fact that realistic potentials are, as a
rule, strongly singular~\cite{classif}.  For a dilute Bose gas this
method, in its simplest form, is reduced to the replacement of the
real potential by the zero-momentum $t$ matrix obtained from the
ordinary two-body problem. In so doing, a new difficulty arises.
Namely, the so-called ultraviolet divergence appears in the
perturbation series.  This feature is usually considered as a
nonfundamental one, and textbooks do not draw much attention to it.
However, some troubles compel us to re-investigate this problem in
more detail.

It is well-known that the first microscopical treatment of the
Bose-Einstein condensation in an interacting Bose gas has been
realized by Bogoliubov in his classical paper of 1947~\cite{Bog1} and
is concerned with the weak-coupling regime. In the Bogoliubov model,
single-particle excitations coincide with the collective ones, the
latter can be obtained within the dielectric formalism in the random
phase approximation, which is reduced to summation of infinite number
of the bubble diagrams~\cite{Noz,Pines}. In order to generalize the
Bogoliubov model to the case of a strongly singular potential one
should take account of the multiple two-particle scattering, and,
thus, summarize infinite number of the ladder [$t$ matrix] diagrams.
The result of this summation is considered to be expressed in the
replacement of the Fourier transform of the pairwise potential
$\Phi(k)$ by the $t$ matrix $t=4\pi\hbar^2a/m$, $a$ being the
scattering length and $m$ denoting the boson mass. However, it is
the replacement that leads to the ultraviolet divergence. This can
easily be shown with the help of the expansion of the energy per
particle for a weakly interacting Bose gas in powers of the gas
parameter $na_0^{3}$ ($n=N/V$ is the boson density):
\begin{equation}
\varepsilon^{(B)}=\frac{2\pi\hbar^2 n(a_{0}+a_{1})}{m}
  +\frac{2\pi\hbar^2 n
        a_{0}}{m}\frac{128}{15\sqrt{\pi}}\sqrt{n a_{0}^3} +\cdots.
\label{eBog}
\end{equation}
Here $a_0$ and $a_1$ are the leading and next-to-leading terms in the
Born series for the scattering length $a$ given by
\begin{eqnarray}
a &=&a_{0}+a_{1}+a_{2}+\cdots, \label{expa}\\
a_{0}&=&\frac{m}{4\pi\hbar^{2}}\Phi(k=0),\;
a_{1}=-\frac{m}{4\pi\hbar^{2}}\int\frac{\d^{3}k}{(2\pi)^{3}}
\frac{\Phi^{2}(k)}{2T_{k}}, \label{aBorn}
\end{eqnarray}
with $T_k=\hbar^2 k^2/(2m)$. For more detail see the article of
Brueckner and Sawada in Ref.~\cite{Lee} and paper~\cite{Ch}. As it is
seen from Eq.~(\ref{aBorn}), the replacement $\Phi(k) \to
4\pi\hbar^2a/m$ yields $a_0 \to a$ and $a_1 \to \infty$. This problem
is considered to be solved since the PhD thesis of Nozi\`eres (1957)
where it has been found that the divergence is an artifact of
combining the bubble and $t$ matrix diagrams~\cite{Pines} and appears
due to double account of the term $2\pi\hbar^2 n a_{1}/m$~($a$
already contains $a_1$!). We emphasize that there is no double
account in the original Bogoliubov approach~($a_0$ does not include
$a_1$). Thus, replacing original interaction $\Phi(k)$ by the $t$
matrix, one should be careful and ignore the divergent term
proportional to $\int\d^{3}k/k^2$. This allows for obtaining the
classical result of Lee and Yang~\cite{LY} given by
\begin{equation}
\varepsilon=\frac{2\pi\hbar^2 n
a}{m}\left(1+\frac{128}{15\sqrt{\pi}}\sqrt{n a^3} + \cdots\right).
\label{eLY}
\end{equation}
and first found in the framework of the binary collision expansion
method. Note that in the pseudopotential formulation escaping the
divergence corresponds to using $V_{eff}=\dfrac{4\pi\hbar^2a}{m}
\dfrac{\partial}{\partial r} r \delta({\bf r})$ instead of
$V_{eff}=\dfrac{4\pi\hbar^2a}{m}\delta({\bf r})$~\cite{Pines}.

The present article shows that the trouble that manifests itself in
the form of the ultraviolet divergence, discussed above, can not be
cured by removing the divergent term. Indeed, in the next section the
kinetic and interaction energies of a strong-coupling Bose gas are
investigated. It is shown that the modified Bogoliubov model, where
the original pairwise potential is replaced by the $t$ matrix, yields
incorrect values of these important quantities. In particular, the
kinetic energy turned out to be negative. Note that the subject of
this section is of interest not only in the particular context of the
effective-interaction scheme. To the best of our knowledge, the
interaction and kinetic energies of a strongly interacting Bose gas
have never been investigated in detail. Moreover, one can find
absolutely different points of view concerning these quantities in
literature. For instance, there is opinion that the mean energy of a
Bose gas, taken in the leading order in the gas parameter $n a^3$, is
equal to the interaction one~\cite{Kett}. According to another point
of view~\cite{Lieb}, all the mean energy of a Bose gas of hard
spheres is kinetic in the same order. Now there is essential need of
clarifying this ambiguous situation because it directly influences
interpretation of the experimental data~(see the first paper in
Ref.~\cite{Kett}). Further, in Sec.~\ref{pairdistr} the pair
distribution function $g(r)$ is calculated within the
effective-interaction approach. This quantity turned out to be
unphysically negative at small boson separations. Remind that
$g(|{\bf r}_1-{\bf r}_2|)/V$ is the density of the conditional
probability of finding one particle at the point ${\bf r}_1$ while
another is at the point ${\bf r}_2$. In Sec.~\ref{pairwaves}
a correct expression for $g(r)$ is derived with the help of the
Hellmann-Feynman theorem and a variational theorem for the scattering
amplitude. To reveal a reason of the failure of the effective-interaction
scheme, we consider a representation for $g(r)$ in terms of
the in-medium pair wave functions. On
the basis of this formula one can conclude that the Bogoliubov model
modified by the replacement $\Phi(k) \to 4\pi\hbar^2a/m$ does not fit
the strong-coupling regime due to weak-coupling
features  that survive in the effective-interaction approach. A
possible way of the strong-coupling generalization of the Bogoliubov
model, which is free from the troubles of the effective-interaction
scheme, is considered in Secs.~\ref{strgeneral}, \ref{short} and
\ref{lowden}. It reproduces the result of Lee and Yang~(\ref{eLY})
and, in addition, yields correct results for the kinetic and
interaction energies and pair distribution function. Note that we
deal with a systematic calculations in the first two orders of the
expansion in $n a^3$. Calculations of the next terms imply additional
and rather extended investigations beyond the scope of the present
paper. By the moment, the authors are able to contemplate some
important points of studying the strong-coupling regime only in terms
of the in-medium pair wave functions. How to proceed with this
interesting problem in the framework of the more familiar Green's
function method is an open question.

\section{The interaction and kinetic energies}
\label{kinint}

Let us start from the original Bogoliubov model and find the
expansions of the kinetic and interaction energies in powers of the
weak-coupling gas parameter $n a^3_0$. The most simple way of doing
this is based on the well-known Hellmann-Feynman theorem given by
\begin{equation}
\delta E=\langle\delta\hat{H}\rangle.
\label{HF}
\end{equation}
Here $\langle\cdots\rangle=\langle0|\cdots|0\rangle$ stands for the ground-state average,
$\delta E$ and $\delta\hat{H}$ are infinitesimal
changes of the ground-state energy
$E=\langle\hat{H}\rangle$ and the Hamiltonian
\begin{equation}
\hat{H}=-\sum_{i}\frac{\hbar^{2}\nabla_{i}^{2}}{2m}
+\frac{1}{2}\sum_{i\not=j}\gamma\Phi(|{\bf r}_{i}-{\bf r}_{j}|),
\label{H}
\end{equation}
where $\gamma$ is an auxiliary parameter, the coupling constant,
which should be put equal to unity in final formulae.
According to this theorem for the kinetic and interaction energies
per particle we have
\begin{eqnarray}
\varepsilon_{int}
&=&\frac{n}{2}\int\d^{3}r\,\gamma\Phi(r)g(r)
=\gamma\frac{\partial\varepsilon}{\partial\gamma},
\label{eint1}\\
\varepsilon_{kin}
&=&\int\frac{\d^{3}k}{(2\pi)^{3}}T_{k}\frac{n_{k}}{n}
=-m\frac{\partial\varepsilon}{\partial m},
\label{HFik}
\end{eqnarray}
where $\varepsilon=E/N=\varepsilon_{kin}+\varepsilon_{int}$ is the total
energy per particle, $n_{k}$ denotes the occupation number. From
Eqs.~(\ref{eBog}), (\ref{aBorn}), (\ref{eint1}) and (\ref{HFik}) one can
derive for the original Bogoliubov~(B) model the following equations:
\begin{eqnarray}
&&\varepsilon^{(B)}_{int}=\frac{2\pi\hbar^2n}{m}
\left(a_{0}+2a_{1}+a_{0}\frac{64}{3\sqrt{\pi}}\sqrt{n a_{0}^{3}}+
\cdots\right),
\label{eintBog}\\
&&\varepsilon^{(B)}_{kin}=\frac{2\pi\hbar^2 n}{m}
\left(-a_{1}-a_{0}\frac{64}{5\sqrt{\pi}}\sqrt{n a_{0}^3}
+ \cdots \right).
\label{ekinBog}
\end{eqnarray}
We remark that $a_1<0$ and, therefore, $\varepsilon_{kin}$
in Eq.~(\ref{ekinBog}) is positive. One can see that
$\varepsilon_{kin} \ll \varepsilon_{int}\;(|a_1| \ll a_0$!) and, thus,
the main part of the mean energy comes from the boson-boson
interaction in the weak-coupling regime.
It should be stressed that the formulae (\ref{eintBog}) and (\ref{ekinBog})
can be obtained directly from the Bogoliubov expressions for $n_{k}$ and $g(r)$,
since the original Bogoliubov model is fully consistent.
Now what we have for the
modified Bogoliubov model with the original interaction $\Phi(k)$
replaced by $t=4\pi\hbar^2a/m$? Substituting $a$ for $a_0$ and
removing the terms depending on $a_1$, one can readily obtain
for the modified Bogoliubov approach~(mB) the following relations:
\begin{eqnarray}
&&\varepsilon^{(mB)}_{int}=\frac{2\pi\hbar^2 a n}{m}
\left(1+\frac{64}{3\sqrt{\pi}}\sqrt{n a^{3}}+
\cdots\right),
\label{eintEfInt}\\
&&\varepsilon^{(mB)}_{kin}=-\frac{2\pi\hbar^2 a n}{m}
\frac{64}{5\sqrt{\pi}}\sqrt{n a^3}+ \cdots
\label{ekinEfInt}
\end{eqnarray}
It is known that $a$ should be positive, otherwise the Bose gas would
be unstable. In this case $\varepsilon^{(mB)}_{kin}$ in
Eq.~(\ref{ekinEfInt}) is negative. Thus, the result given by
Eqs.~(\ref{eintEfInt}) and (\ref{ekinEfInt}) can not be correct.

What are the true values of $\varepsilon_{kin}$ and $\varepsilon_{int}$
in the strong-coupling regime? This can again be clarified with the help
of the Hellmann-Feynman theorem (\ref{HF}). However, first of all
one should prove a useful variational theorem:
\begin{eqnarray}
\delta{U}^{(0)}(0)=\int\d^3r
\Bigl[
&&\psi^{(0)}(r)\delta\Bigl(-\frac{\hbar^2}{m}\nabla^2 \Bigr)\psi^{(0)}(r)
\nonumber \\
&&+\varphi^{(0)}(r)\delta\bigl(\Phi(r)\bigr)\varphi^{(0)}(r)
\Bigr],
\label{delU0}
\end{eqnarray}
where we define the scattering amplitude by the relation
\begin{equation}
U^{(0)}(k)=\int\d^3r\,\varphi^{(0)}(r)\Phi(r) \exp(i{\bf k}{\bf r}),
\label{U0}
\end{equation}
and $\varphi^{(0)}(r)=1+\psi^{(0)}(r)$ is the solution of the two-body
Schr\"odinger equation in the centre-of-mass system
\begin{equation}
-\frac{\hbar^2}{m}\nabla^2\varphi^{(0)}(r)+\Phi(r)\varphi^{(0)}(r)=0
\label{twobody}
\end{equation}
with the asymptotic behaviour at $r \to \infty$
\begin{equation}
\varphi^{(0)}(r)\to 1-a/r.
\label{bound}
\end{equation}
From Eqs.~(\ref{twobody}) and (\ref{bound}) it follows that $U^{(0)}(0)=
4\pi\hbar^2a/m$. Relation (\ref{delU0}) can be proved representing
Eq.~(\ref{U0}) in the form
\begin{eqnarray}
{U}^{(0)}(0)
=\int\d^3r\Bigl[\frac{\hbar^2}{m}|\nabla\psi^{(0)}(r)|^2
                            +[\varphi^{(0)}(r)]^2 \Phi(r)\Bigr],
\label{U01}
\end{eqnarray}
using integration by parts, and taking into account the
Schr\"odinger equation (\ref{twobody}) and the boundary condition
(\ref{bound}). Further, varying  Eq.~(\ref{U01}), we get
Eq.~(\ref{delU0}) with Eqs.~(\ref{twobody}) and (\ref{bound}) taken
into account.

From the theorem (\ref{delU0}) it follows that
\begin{equation}
\gamma\frac{\partial a}{\partial \gamma}=
m\frac{\partial a}{\partial m}=a-b
=\frac{m}{4\pi\hbar^2}\int\d^3r\,[\varphi^{(0)}(r)]^2\Phi(r),
\label{da}
\end{equation}
where we introduce by definition the quantity
\begin{equation}
b=\frac{1}{4\pi}\int\d^{3}r\,\bigl|\nabla\psi^{(0)}(r)\bigr|^{2}.
\label{b}
\end{equation}
As it is seen, Eqs.~(\ref{eLY}), (\ref{HFik}), (\ref{eint1}) and (\ref{da}) provide
the following expressions:
\begin{eqnarray}
&&\varepsilon_{int}=\frac{2\pi \hbar^2 (a-b)n}{m}
\left(1+\frac{64}{3\sqrt{\pi}} \sqrt{n a^3}+ \cdots \right),
\label{eint}\\
&&\varepsilon_{kin}=\frac{2\pi \hbar^2 b n}{m}
\left(1+\frac{64}{3\sqrt{\pi}}\sqrt{n a^3}
\left(1-\frac{3a}{5b}\right)+ \cdots \right).
\label{ekin}
\end{eqnarray}
Equation~(\ref{b}) implies that $b$ is a positive quantity that can be
considered as a new characteristic length. We stress that $b$ is expressed in
terms of $a$ and its derivatives [see Eq.~(\ref{da})] and determined by a
shape of the interaction potential. For example, when $\Phi(r)$ is the
hard-sphere potential
\begin{equation}
\Phi(r)=\left\{\begin{array}{ll}
+\infty, & r<a \\
0,       & r>a,
\end{array}\right.
\label{hards}
\end{equation}
we have $b=a$, which follows from the obvious relations $\partial a/\partial
\gamma=\partial a/\partial m=0$. While for the weak-coupling
potential~\cite{classif} $b\simeq -a_{1}$, $a \simeq a_{0}$, and, hence, $b
\ll a$. The latter implies that $\varepsilon_{int} \approx \varepsilon$ and
$\varepsilon_{kin} \approx 0$ in the weak-coupling regime.  This allows for
concluding that $\varepsilon_{int}$ and $\varepsilon_{kin}$ depends on a
particular shape of the interaction potential even in the leading order of
the expansion in $n a^3$.  Equations (\ref{eint}) and (\ref{ekin}) testify
that in the case of the hard-sphere interaction (\ref{hards}) all the mean
energy is kinetic, which agrees with the expectation of Lieb and
Yngvason~\cite{Lieb} and definitely excludes the variant adopted in
Ref.~\cite{Kett}.  One can see that this result is rather general: {\it for
the hard-sphere potential (\ref{hards}) the interaction energy is equal to
zero for any density.} Indeed, $\Phi(r)$ given by (\ref{hards}) can be
thought of as the limiting case of the repulsive potential
\[
\Phi(r)=\left\{\begin{array}{ll}
V_{0},   & r<a \\
0,       & r>a.
\end{array}\right.
\]
It is clear that saturation takes place when $V_{0}\gg \varepsilon$:
further increase of the parameter $V_{0}$ does not change the energy per
particle $\varepsilon$. Hence, according to Eq.~(\ref{eint1}),
$\varepsilon_{int}=0$ because $\partial\varepsilon/\partial \gamma=0$ at
$\gamma=1$ in the limit $V_{0}\to +\infty$.  Thus, the incorrect results of
the effective-interaction scheme given by Eqs.~(\ref{eintEfInt}) and
(\ref{ekinEfInt}) are, say, of the weak-coupling character
as they imply the relations $\varepsilon_{int} \approx \varepsilon$ and
$\varepsilon_{kin}\approx 0$ in the leading order.

\section{Pair distribution in the effective-interaction approach}
\label{pairdistr}

It turned out that the trouble concerning the interaction and kinetic
energy is accompanied within the effective-interaction approach by
the problem related to the pair spatial correlations. Let us show
this with the help of the Beliaev's paper and article of Hugenholtz
and Pines. In the latter the structural factor $S(k)$ has been
calculated. We can employ it to find the pair distribution function
$g(r)$ with the well-known relation:
\begin{equation}
g(r)=1+\frac{1}{n} \int \frac{\d^3k}{(2\pi)^3} \left[ S(k)- 1\right]
\exp(i{\bf k}{\bf r}).
\label{g}
\end{equation}
However, Hugenholtz and Pines~\cite{Lee} restricted themselves to the approximation
valid at $k \to 0$, which makes it impossible to integrate their result to
get $g(r)$. So, their calculations are repeated below but not involving the
argument of $k \to 0$. They investigated $S(k)$ with the help of the
Green function
\begin{equation}
F_k(t-t')=-\frac{i}{\hbar}\langle0|T\{\hat{\rho}_{{\bf k}}(t)
 \hat{\rho}_{-{\bf k}}(t')\}|0\rangle,
\label{F}
\end{equation}
where $\hat{\rho}_{\bf k}=\sum_{{\bf k}} \hat{a}^{\dagger}_{\bf q}
\hat{a}_{{\bf q}+{\bf k}}$ is the Fourier transform of the density
operator. According to the paper of Hugenholtz and Pines~\cite{Lee}, in the
lowest order in the condensate depletion $(n-n_0)/n$
(when one can take $n=n_0$, $n_{0}$ is the density of the condensate) the quantity $F_k(t-t')$ can be written as
\begin{eqnarray}
&&F_k(t-t')\nonumber\\
&&=-\frac{i}{\hbar}
    N\langle 0|T\{\hat{a}^{\dagger}_{-\bf k}(t)\!\!+
                                    \!\!\hat{a}_{\bf k}(t)\}
\{\hat{a}^{\dagger}_{\bf k}(t')\!\!+\!\!\hat{a}_{-\bf k}(t')\}|0\rangle,
\label{F1}
\end{eqnarray}
so that its Fourier transform is given by
\begin{eqnarray}
&&F(k,\omega)=\nonumber \\
&&\frac{N\left[2T_k+\Sigma^{+}_{11}+\Sigma^{-}_{11}-2\mu-
2\Sigma_{02} \right]}{\left[\hbar\omega\!-\!\frac{1}{2}(\Sigma^{+}_{11}\!-
\!\Sigma^{-}_{11})\right]^2\!\!\!-
  \!\left[T_k\!-\!\mu\!+\!\frac{1}{2}(\Sigma^{+}_{11}\!+
\!\Sigma^{-}_{11})\right]^2\!\!\!+\!\Sigma^2_{02}\!+\!i\delta},
\nonumber \\
\label{FourF}
\end{eqnarray}
where $\Sigma^{+}_{11},\Sigma^{-}_{11}$ and $\Sigma_{02}$ are the
effective potentials introduced by Beliaev; $\mu$ stands
for the chemical potential. Beliaev has found that in the lowest order
in the condensate depletion the effective potentials and $\mu$ obey
the following relations:
\begin{equation}
\mu =nf(0,0),\;\Sigma_{02}=nf({\bf k},0),\;\Sigma^{\pm}_{11}=
2nf_s({\bf k}/2,{\bf k}/2).
\label{effpot}
\end{equation}
Here the expression $f({\bf k},{\bf p})$ is the ``non-diagonal"
scattering amplitude:
\begin{equation}
f({\bf k},{\bf p})=\int\d^3r\varphi^{(0)}_{{\bf p}}({\bf r})\Phi(r)
\exp(-i{\bf k}{\bf r}),
\label{fk0}
\end{equation}
where $\varphi^{(0)}_{{\bf p}}({\bf r})$ obeys the Schr\"odinger equation
(\ref{twobody}) with the right-hand side equal to
$(\hbar^{2}p^{2}/m)\varphi^{(0)}_{{\bf p}}({\bf r})$~\cite{NoteB}.
Besides, $f_s({\bf k},{\bf k'})=[f({\bf k},{\bf k'})+f(-{\bf k},{\bf k'})]
/2$. Substituting Eq.~(\ref{effpot}) in Eq.~(\ref{FourF}), we arrive at
\begin{eqnarray}
F(k,\omega)=\frac{2N
 \bigl[T_k\!+\!n\bigl(2f_s({\bf k}/2,{\bf k}/2)\!-\!f({\bf k},0)\!-\!
f(0,0)\bigr)\bigr]}{\hbar^2\left(\omega -\omega_k + i\delta\right)
\left(\omega+\omega_k-i\delta\right)}
\nonumber
\end{eqnarray}
with
$$
\hbar\omega_k
      =\sqrt{\bigl(T_k\! +2nf_s({\bf k}/2,{\bf k}/2)-nf(0,0)\bigr)^2-
n^2f^2({\bf k},0)}.
$$
Now, to derive the structural factor $S(k)$, one should utilize
the definition~(\ref{F}) at $t=t'$:
\[
S(k)=\frac{\langle\hat{\rho}_{{\bf k}}\hat{\rho}_{-{\bf k}}\rangle}{N}
=\frac{1}{N} \int\limits_{-\infty}^{+\infty}\d\omega\frac{i\hbar}{2\pi}
F(k,\omega).
\]
After transparent integration one can find
\begin{equation}
S(k)=\frac{T_k+n\bigl(2f_s({\bf k}/2,{\bf k}/2)-f({\bf k},0)-
f(0,0)\bigr)}{\hbar\omega_k}.
\label{str1}
\end{equation}
With the help of the limiting relation at $k \to 0$
$$
f_s({\bf k}/2,{\bf k}/2)\simeq f({\bf k},0) \simeq f(0,0)=
\frac{4\pi\hbar^2a}{m}
$$
equation~(\ref{str1}) is reduced to the familiar expression
\begin{equation}
S(k)\simeq
\frac{T_k}{\hbar\omega_k}\simeq\frac{T_k}{\sqrt{T^2_k+2nT_kf(0,0)}}=
\frac{k^2}{\sqrt{k^4+16\pi n a k^2}}.
\label{str2}
\end{equation}
As it has been mentioned before, Eq.~(\ref{str2}) can not be used in
Eq.~(\ref{g}) because the integral diverges in this situation at large
momenta. On the contrary, the exact (in the lowest order in the depletion)
variant (\ref{str1}) can be employed without problems concerning
integration. Inserting Eq.~(\ref{str1}) in Eq.~(\ref{g}), at $n \to 0$
we get
\begin{equation}
g(r) \to 1-\int \frac{\d^3k}{(2\pi)^3} \frac{f({\bf k},0)}{T_k}
\exp(i{\bf k}{\bf r}).
\label{g1}
\end{equation}
Equations~(\ref{twobody}) and (\ref{fk0}) at ${\bf p}=0$ allow for
representing Eq.~(\ref{g1}) in the limit $n \to 0$ as
\begin{equation}
\lim_{n\to0}g(r)=1+2\psi^{(0)}(r).
\label{g2}
\end{equation}
Now we can easily be convinced that the result for $g(r)$ derived within the
effective-interaction scheme is inadequate in the strong-coupling
regime~\cite{classif}.  Indeed, in this regime $\varphi^{(0)}(r=0)=0$ and, by
definition, $\psi^{(0)}(r=0)=-1$. This, taken together with Eq.~(\ref{g2}),
leads to $g(r=0) \to -1$ when $n \to 0$. Thus, the pair distribution function
inferred from Eq.~(\ref{str1}) is negative at small boson separations, which
implies unphysical picture of the short-range boson correlations.

In addition, one can see from Eq.~(\ref{g2}) that the effective-interaction
approach is not self-consistent. Indeed, in the case of a strongly singular
potential we have $g(r=0) \to -1$ at $n \to 0$. Then, from the first equality
in Eq.~(\ref{eint1}) and Eq.~(\ref{g2}) one can find
$\varepsilon^{(mB)}_{int}$ to be infinite~(one more divergence!). However,
Eq.~(\ref{eintEfInt}) gives the finite value for $\varepsilon^{(mB)}_{int}$.
Thus, two different ways of calculating the interaction energy yield two
different results.

\section{Pair wave functions}
\label{pairwaves}

\begin{figure}[tb]
\epsfxsize=65mm \centerline{\epsfbox{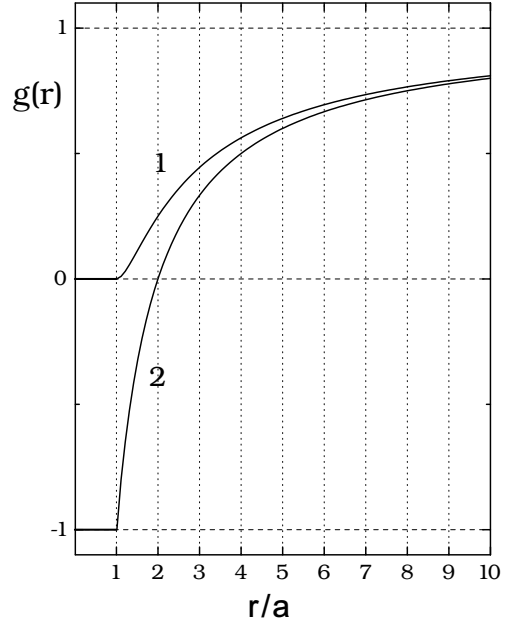}}
\vspace{2mm}
\caption{The pair distribution function for the hard-sphere
bosons in the limit $n \to 0$:
1 - the data of Eq.~(\ref{gcorr2}); 2 - the result of the
effective-interaction approach~\protect\cite{Lee}, Eq.~(\ref{g2}).
At $r<2a$ Eq.~(\ref{g2}) yields unphysically negative
values for $g(r)$.}
\label{fig1}
\end{figure}

Now the question arises: what behaviour is correct for $g(r)$ in a dilute
strongly interacting Bose gas? This can be clarified with the help of the
theorems (\ref{HF}) and (\ref{delU0}), starting from the correct
expansion~(\ref{eLY}).  For a homogeneous system the pair distribution
function can be represented as
\[
g(r)=\frac{V}{N(N-1)}
\left\langle\sum_{i\not=j}\delta({\bf r}-{\bf r}_{i}+{\bf r}_{j})
                                                      \right\rangle.
\]
Here $\delta({\bf r})$ stands for the Dirac's delta-function.
The Hellman-Feynman theorem (\ref{HF}) yields
\[
g(r)=\frac{2}{n}\frac{\delta\varepsilon}{\delta\Phi(r)},
\]
where we put $(N-1)/V\simeq n$ in the thermodynamic limit.
Substituting the expression (\ref{eLY}) and using the theorem
(\ref{delU0}), we arrive at
\begin{equation}
g(r)=[\varphi^{(0)}(r)]^2
\left(1+\frac{64}{3\sqrt{\pi}}\sqrt{na^3}+\cdots\right).
\label{grnto01}
\end{equation}
This equation is discussed in detail in Sec.~\ref{short}. In the limit
$n\to0$ from Eq.~(\ref{grnto01}) we obtain the expression known
since the Bogoliubov's article [see the concluding part of his classical
paper~\cite{Bog1}, where a possible way of going beyond the Born
approximation is discussed]:
\begin{equation}
\lim_{n\to0}g(r)=\bigl[\varphi^{(0)}(r)\bigr]^2
                =\bigl(1+\psi^{(0)}(r)\bigr)^2.
\label{gcorr2}
\end{equation}
Here $\varphi^{(0)}(r)$ obeys the Schr\"odinger equation (\ref{twobody})
with the boundary condition (\ref{bound}).
To compare the results given by Eqs.~(\ref{g2}) and (\ref{gcorr2}),
let us consider the simplest case of the hard-sphere pairwise
interaction defined by Eq.~(\ref{hards}). In this case we get
\begin{equation}
\varphi^{(0)}(r)=\left\{\begin{array}{ll}
0,       & r\leq a \\
1-a/r,   & r>a.
\end{array}\right.
\label{hardsphi0}
\end{equation}
The data found from Eqs.~(\ref{g2}) and (\ref{gcorr2}) for the
hard-sphere bosons are presented in Fig.~\ref{fig1}.
As it is seen, at small particle separations $r\lesssim 3a$ the
difference between the curves (1) and (2) is essential, while at
large $r$ these curves are close to one another.  We emphasize that
the incorrect behaviour at the short distances is related to the
modified Bogoliubov's model [with the replacement
$\Phi(k)\to2\pi\hbar^{2}a/m$] but not to the original one developed for
the case of a weakly interacting Bose gas.

The failure of the effective-interaction approach can be understood
with the help of the interesting relation connecting the pair
distribution function with the in-medium pair wave functions and
following from the Bogoliubov principle of the correlation
weakening~\cite{Ch1,Ch3,Ch2}. We remind that the pair wave
functions, by definition, are the eigenfunctions of the reduced
density matrix of the second order. For the Bose gas, a system
with a small condensate depletion $(n-n_0)/n \ll 1$, the pair
distribution function can be written as
\begin{equation}
g(r)=\left(\frac{n_0}{n}\right)^2\varphi^2(r)+2\frac{n_0}{n}
\int\frac{\d^3q}{(2\pi)^3}\frac{n_q}{n}\varphi^2_{{\bf q }/2}({\bf r}),
\label{gpw}
\end{equation}
which is accurate to the next-to-leading order in $(n-n_0)/n$~\cite{Note2}.
Another restriction for this representation is an assumption that there are
no bound pair states (see details in Ref.~\cite{Ch1}).  In Eq.~(\ref{gpw})
$\varphi(r)$ is the in-medium pair wave function of two condensed bosons;
$n_q=\langle a_{{\bf q}}^{\dagger}a_{{\bf q}}\rangle$; the quantity
$\varphi_{{\bf q}/2} ({\bf r})$ denotes the in-medium wave function of the
relative motion of the pair of bosons with the total momentum $\hbar{\bf q}$.
This pair includes one condensed and one uncondensed particle. The functions
$\varphi(r)$ and $\varphi_{{\bf q}/2}({\bf r})$ are chosen as real
quantities. It is also convenient to introduce the in-medium scattering waves
$\psi(r)$ and $\psi_{{\bf p}}({\bf r})\;(p\not=0)$ given by
\begin{equation}
\varphi(r)=1+\psi(r),\;
\varphi_{{\bf p}}({\bf r})= \sqrt{2}\cos({\bf p}{\bf r})+
\psi_{{\bf p}}({\bf r})
\label{scw}
\end{equation}
with the boundary conditions at $r \to \infty$
\begin{equation}
\psi(r),\;\psi_{{\bf p}} ({\bf r}) \to 0.
\label{bc}
\end{equation}
The Fourier transforms of the scattering waves can be expressed
in terms of the Bose operators $a_{{\bf p}}^{\dagger}$ and
$a_{{\bf p}}$~\cite{Ch1}:
\begin{eqnarray}
\psi(k)=
  \frac{\langle a_{{\bf k}}a_{-{\bf k}}\rangle}{n_0},\
{\psi}_{\bf p}({\bf k})= \sqrt{\frac{V}{2 n_0}}
\frac{\langle a^{\dagger}_{2{\bf p}}
        a_{{\bf p}+{\bf k}} a_{{\bf p}-{\bf k}}\rangle}{n_{2p}}.
\label{25}
\end{eqnarray}
At $ n \to 0$ the in-medium pair wave functions tend to
solutions of the ordinary two-body problem. For example, $\varphi(r)
\to \varphi^{(0)}(r)$. Hence, as the condensate depletion $(n-n_0)/n$
approaches zero when $n\to 0$, Eq.~(\ref{gpw}) is reduced to
the Bogoliubov's equation (\ref{gcorr2}). A weakly interacting Bose gas is characterized by a
minor role of the particle scattering, so that $|\psi(r)| \ll 1$ and
$|\psi_{{\bf p}}({\bf r})| \ll 1$. In particular, the Bogoliubov
model of a weak-coupling Bose gas implies $\psi_{{\bf p}}({\bf
r})=0$~\cite{Ch1,Ch3}, while $\psi(r)\not=0$. Thus, within the
Bogoliubov model, Eq.~(\ref{gpw}) can be approximated as
\begin{eqnarray}
g(r)=1+2\psi(r)+\frac{2}{n} \int \frac{\d^3k}{(2\pi)^3}
     n_k\exp(i{\bf k}{\bf r}),
\label{gpwB}
\end{eqnarray}
where the terms of the order of $(n-n_0)^2/n^2$ and $\psi(r)(n-n_0)/n$
have been ignored. At $n \to 0$ one can find~\cite{Ch,Ch3}
\begin{equation}
g(r) \to 1+2\psi^{(0)}_B(r),
\label{gcorrB}
\end{equation}
where $\psi^{(0)}_B=\varphi^{(0)}_B(r)-1$ obeys the equation
\begin{equation}
-\frac{\hbar^2}{m}\nabla^2\varphi^{(0)}_B(r)+\Phi(r)=0,
\label{tbB}
\end{equation}
which is the two-body Schr\"odinger equation (\ref{twobody})
taken in the Born approximation. We stress that
Eqs.~(\ref{gpwB})-(\ref{tbB}) are related to the original Bogoliubov
model. As to the modified Bogoliubov approach with the $t$ matrix, it
produces Eq.~(\ref{g2}). Now we are able to explain the failure of
the effective-interaction scheme. {\it The replacement of the
original interaction $\Phi(k)$ by the $t$ matrix leads to passage
from $\psi^{(0)}_B(r)$ to $\psi^{(0)}(r)$. However, it does not
influence the Bogoliubov ansatz (\ref{gpwB}) for the relation between
the pair distribution function and the pair wave functions.}
Substituting $\psi(k)=\langle a_{{\bf k}}a_{-{\bf
k}}\rangle/n_0$ and $n_k= \langle a^{\dagger}_{{\bf k}}a_{{\bf
k}}\rangle$ in Eq.~(\ref{gpwB}) and using Eq.~(\ref{g}), one can
see that the approximation (\ref{gpwB}) is consistent with the approximation
(\ref{F1}) for the Green function (\ref{F}).
Thus, the effective-interaction scheme combines the features of both the
strong- and weak-coupling regimes, and this is the actual reason for its
problems discussed in the previous sections. In particular, the ultraviolet
divergence does come from the double account mentioned in Sec.~\ref{1sec},
but our analysis reveals that the double account, in turn, is a consequence
of the combination just pointed out. Note that the thermodynamic
inconsistency of the modified Bogoliubov model, found in our previous
publications~\cite{Ch,Ch3}, results from this combination, too.

\section{Strong-coupling generalization of the Bogoliubov model}
\label{strgeneral}

It follows from the consideration given above that in the
strong-coupling regime one should leave the effective-interaction
approach and develop a new one. What way should one prefer? At
present the authors have no final solution providing the solid
theoretical scheme and making it possible to realize systematic
calculations like in the weak-coupling case. However, we can propose
a reasonable strong-coupling generalization of the Bogoliubov model
based on Eq.~(\ref{gpw}) and semi-phenomenological relation
(\ref{mdscw}) discussed below. This generalization is justified as it
reproduces the result of Lee and Yang (\ref{eLY}), gives correct
picture of the short-range particle correlations, and yields
Eqs.~(\ref{eint}) and (\ref{ekin}). Besides, the way proposed
provides self-consistent calculations of the in-medium pair wave
functions approaching solutions of the ordinary two-body problem at
$n \to 0$. The model considered below is similar to the Brueckner
theory~(see Secs. 36 and 41 in Ref.~\cite{Fet}) but with one
advantageous exception. The point is that the Brueckner theory
implies the momentum distribution of interacting particles to be the
same as in an ideal gas. As to our model, it results in a system
of equations connecting the in-medium pair wave functions with the
momentum distribution.

In addition to Eq.~(\ref{gpw}), the proposed strong-coupling
generalization of the Bogoliubov model involves two other relations.
The first is the familiar expression for the mean energy per particle
\begin{equation}
\varepsilon=\int \frac{\d^3k}{(2\pi)^3} T_k \frac{n_k}{n}
            +\frac{n}{2}\int \d^3r\,g(r)\Phi(r),
\label{emean}
\end{equation}
which can be found in any textbook on statistical mechanics. The
second is, say, the semi-phenomenological relation that connects the
scattering waves (\ref{scw}) with the momentum distribution $n_k$:
\begin{equation}
n_k(n_k+1)= n_0^2{\psi}^2(k)+ 2n_0\int\frac{\d^3q}{(2\pi)^3}n_q
                                      {\psi}^2_{{\bf q}/2}({\bf k}).
\label{mdscw}
\end{equation}
It is worth saying several words about Eq.~(\ref{mdscw}). This
equation represents, in particular, the well-known fact that when
there is no scattering [interaction] in the system, there are no
uncondensed bosons $n_k=0$~\cite{Note3}. The larger interaction, the
larger depletion of the Bose condensate. Equation~(\ref{mdscw})
generalizes the similar relation of the Bogoliubov model given
by~(see Refs.~\cite{Ch,Ch3})
\begin{equation}
n_k(n_k+1) = n^2{\psi}^2(k).
\label{mdscwB}
\end{equation}
The generalization (\ref{mdscw}) has been chosen in \cite{Ch,Ch3} for
the following reasons. First, it provides the relation
$\sqrt{2}\psi(r)=\lim_{p \to 0} \psi_{{\bf p}} ({\bf r})$ as well as
\begin{equation}
\sqrt{2}\varphi(r)=\lim_{p \to 0} \varphi_{{\bf p}} ({\bf r}),
\label{phiphi0}
\end{equation}
which can be inferred from Eqs.~(\ref{scw}) and (\ref{bc}). Second,
it leads to the correct thermodynamics and behaviour of $g(r)$ at
short distances for a dilute Bose gas in the leading and
next-to-leading orders in $n a^3$. Third, the relation (\ref{mdscw})
results in  Eq.~(\ref{LSch}) that gives not only short-range but
also correct long-range behaviour of $\varphi(r)$.  A shortcoming of
this generalization is that it may be not unique, but it is the
simplest one that provides the points mentioned above.

Equations~(\ref{gpw}) and (\ref{emean}) make it possible to express
$\varepsilon$ in terms of the pair wave functions and momentum
distribution. So, a variational procedure can be employed to determine
these quantities. Perturbing ${\psi}(k)$ and $n_k$ and bearing in mind
(\ref{mdscw}), from Eqs.~(\ref{gpw}) and (\ref{emean}) we find the
following equation~\cite{Ch3}:
\begin{equation}
-2\widetilde{T}_k {\psi}(k)
={U}(k)\bigl[1+2\bigl(n_k+n{\psi}(k)\bigr)\bigr],
\label{BG}
\end{equation}
where $\psi(k)$ and $\psi_{{\bf p}}({\bf k})$ are the Fourier transforms
of the scattering waves (\ref{scw}). Here ${U}(k)$ and $\widetilde{T}_k$
are defined by
\begin{eqnarray}
{U}(k)&=&\int\d^3r\,\varphi(r)\Phi(r)\exp(-i{\bf k}{\bf r}),
\label{U}\\
\widetilde{T}_k&=&T_k+nt_{k},
\label{wTk}
\end{eqnarray}
where $t_{k}=U'(k)-U(k)$ and
\begin{eqnarray}
U'(k)&=&
\int\d^3r\,\bigl(\varphi_{{\bf k}/2}^2({\bf r})-\varphi^2(r)\bigr)\Phi(r)
\nonumber \\
&&-\int\frac{\d^3q}{(2\pi)^3}\frac{{U}(q)
   \bigl({\psi}^2_{{\bf k}/2}({\bf q})-{\psi}^2(q)\bigr)}{{\psi}(q)}.
\label{U'}
\end{eqnarray}
Note that $t_k \propto k^4$ for $k \to 0$~\cite{Ch3}, and
$\widetilde{T}_k/T_k \to 1$ when $k \to \infty$. So, at low momenta
and small densities we have $n t_k\ll T_k$. Owing to this property
the difference between $T_k$ and $\widetilde{T}_k$ does not play a
role when calculating the first two terms of the low-density
expansions for the basic thermodynamic quantities. Therefore, at
sufficiently small values of $n$ Eq.~(\ref{BG}) is reduced to
\begin{eqnarray}
\frac{\hbar^2}{m}\nabla^2\varphi(r)&=&\varphi(r)\Phi(r)\nonumber \\
&&+n\int\d^3y\,\varphi(y)\Phi(y)\bigl(g_{tr}(|{\bf r}-{\bf y}|)-1\bigr),
\label{BG1}
\end{eqnarray}
where $g_{tr}(r)$ stands for the truncated pair distribution function
that is equal to the right-hand side of Eq.~(\ref{gpwB}) even beyond
the weak-coupling regime. Equation~(\ref{BG1}) is very similar to the
Bethe-Goldstone equation~\cite{Bethe}. However, there is also
difference between them. Indeed, $g_{tr}$ depends not only on the
scattering wave $\psi(k)$ but on the momentum distribution $n_k$,
too. While the Bethe-Goldstone equation involves one unknown, the pair
wave function.

Equations (\ref{gpw}) and (\ref{mdscw}) are accurate to the next-to-leading
order in $(n-n_0)/n$, then, Eq.~(\ref{BG}) can be accurate only to the
leading order in $(n-n_0)/n$. So, to find $\psi(k)$ and $n_k$, one should
solve Eq.~(\ref{BG}) in conjunction with Eq.~(\ref{mdscw}) taken in the
approximation valid to the leading order in the condensate depletion.
In other words, one has to consider the system of Eqs.~(\ref{mdscwB}) and
(\ref{BG}) which has the following solution:
\begin{eqnarray}
    n_k&=&
\frac{1}{2}\Biggl(\frac{\widetilde{T}_k+nU(k)}
{\sqrt{\widetilde{T}_k^2+2n\widetilde{T}_k U(k)}}
-1\Biggr),
\label{md} \\
\psi(k)&=&
-\frac{1}{2}\frac{U(k)}{\sqrt{\widetilde{T}_k^2+2n\widetilde{T}_k
U(k)}}.
\label{psi}
\end{eqnarray}
Note that as to the scattering states of a pair made of one
condensed and one uncondensed boson, the goal of the present
investigation makes it possible not to go into details. Below it is
only sufficient to limit ourselves to the relation~(\ref{phiphi0}).

In the zero-density limit, Eq.~(\ref{psi}) is reduced to ${\psi}(k)=
\psi^{(0)}(k)= -{U}^{(0)}(k) /(2T_k)$, which can be rewritten in the
form of Eq.~(\ref{twobody}). So, at sufficiently small densities we
can express the quantities ${\psi}(k)$ and $n_k$ in terms of the
vacuum scattering amplitude ${U}^{(0)}(k)$ given by
Eq.~(\ref{U0}) [in the Beliaev's notations ${U}^{(0)}(k)=f({\bf
k},0)$]. This is totally consistent with the well-known argument of
Landau~\cite{Note4} according to which the thermodynamics of dilute
quantum gases is determined by the vacuum scattering amplitude.
Note that the expressions for ${\psi}(k)$ and $n_k$ derived within the
original Bogoliubov model can be obtained~\cite{Ch,Ch3} from
Eqs.~(\ref{md}) and (\ref{psi}) with replacement of $\widetilde{T}_k$ and
${U}(k)$ by $T_k$ and ${\Phi}(k)$, respectively. So, in what
concerns the expressions for $n_k$ and $\psi(k)$, the situation
looks as if we operated with a Bose gas of weakly interacting
quasiparticles with the renormalized kinetic energy $\widetilde{T}_k$
and effective interaction $U(r)=\varphi(r)\Phi(r)$. This is indeed close
to the expectations based on the effective-interaction approach of the
papers~\cite{Lee}.

\section{Short-range boson spatial correlations}
\label{short}

Now, to elaborate on the picture of the short-range boson
correlations, let us investigate how the correlation hole stipulated
by the repulsion between bosons at small separations changes under
the influence of the surrounding bosons. At $n \to 0$ this hole is
completely specified by the condensate-condensate pair wave function
$\varphi(r)$, which can be found from the definition~(\ref{U}).
Using this definition and Eq.~(\ref{psi}), for the
scattering amplitude one can find
\begin{equation}
{U}(k)={\Phi}(k)-
      \frac{1}{2}\int\frac{\d^3q}{(2\pi)^3}\frac{{\Phi}(|{\bf k}
                  -{\bf q}|){U}(q)}{\sqrt{\widetilde{T}^2_q
                                +2n\widetilde{T}_q{U}(q)}},
\label{LSch}
\end{equation}
which is the in-medium Lippmann-Schwinger equation. Let us
rewrite Eq.~(\ref{LSch}) in the form
$$
{U}(k)={\Phi}(k)-\frac{1}{2}\int\frac{\d^3q}{(2\pi)^3}
                 \frac{{\Phi}(|{\bf k}-{\bf q}|){U}(q)}{T_q}-I,
$$
where for $I$ we have
$$
I=\frac{1}{2}\int\frac{\d^3q}{(2\pi)^3}\Biggl[\frac{{\Phi}(|{\bf k}
  -{\bf q}|){U}(q)}{\sqrt{\widetilde{T}^2_q+ 2n\widetilde{T}_q{U}(q)}}-
  \frac{\Phi(|{\bf k}-{\bf q}|){U}(q)}{T_q}\Biggr].
$$
Performing the ``scaling" substitution
\begin{equation}
{\bf q}={\bf q'}\sqrt{2mn}/\hbar
\label{subst}
\end{equation}
in the integral and, then, taking the zero-density limit
in the integrand, for $n \to 0$ we find~\cite{Note5}
\begin{equation}
I= -\alpha {\Phi}(k),
\quad \alpha=\frac{\sqrt{n m^3}}{\pi^2\hbar^3}{U}^{3/2}(0).
\label{In0}
\end{equation}
From Eqs.~(\ref{LSch}) and (\ref{In0}) it now follows that
\begin{eqnarray}
U(k)-{U}&&^{(0)}(k)=\alpha{\Phi}(k)\nonumber \\
&&-\int \frac{\d^3q}{(2\pi)^3} \frac{{\Phi}(|{\bf k}
   -{\bf q}|)}{2T_q}\bigl[{U}(q)
                -{U}^{(0)}(q)\bigr],
\label{LSch1}
\end{eqnarray}
where $U^{(0)}(k)$ obeys Eq.~(\ref{LSch}) with $n=0$, i.e. the standard
Lippmann-Schwinger equation. Introducing the quantity
${\xi}(q)=-[{U}(q) -{U}^{(0)}(q)]/(2T_q)$, for its Fourier transform
$\xi(r)$ we find the equation that is nothing else but
the Schr\"odinger equation (\ref{twobody}) with $\varphi^{(0)}(r)$ replaced by
$\alpha+\xi(r)$. As $\xi(r) \to 0$ when $r \to \infty$, we can
conclude that $\xi(r)=\alpha \psi^{(0)}(r)$. Hence, for $n \to 0$ we get
\begin{equation}
{U}(k) \simeq {U}^{(0)}(k)\left(1 +\gamma(k,n)
\frac{8}{\sqrt{\pi}}\sqrt{na^3}\,\right).
\label{UkU0k}
\end{equation}
Here $\gamma(k,n) \to 1$ when $n\to 0$. At $k=0$ the derived result
for, say, the in-medium scattering amplitude $U(k)$ coincides with
the expansion in $n a^3$ for the effective potential found within the
effective-interaction approach at the zero temperature (see,
e.g., the review \cite{ShiGrif}, Eq.~(4.27)). This shows once more that
there are actual parallels between our model and approach of
Ref.~\cite{Lee}. However, these parallels are accompanied by
significant differences. First, the in-medium Lippmann-Schwinger
equation (\ref{LSch}) is not a variant of the $t$ matrix equation
which is frequency dependent, contrary to Eq.~(\ref{LSch}).
Second, Eq.~(\ref{LSch}) has been found beyond any diagram technique
by means of the variational procedure. One of its important
consequences is that the pair wave functions that ``generate" the
in-medium scattering amplitudes [see Eq.~(\ref{U}) and Eq.~(\ref{Up}) below] in our
approach coincide with the pair wave functions that make a
contribution to $g(r)$ [see Eq.~(\ref{gpw})].  By contrast, the
modified Bogoliubov model implies the plane waves for
$\varphi_{{\bf p}}({\bf r})$ ($p\not=0$) in the pair distribution
function (\ref{gpwB}) (see Sec.~\ref{pairwaves}), while one certainly
goes beyond the plane-wave approximation when calculating $t$ matrix
corresponding to a pair of particles with nonzero total momenta.

With the help of Eq.~(\ref{UkU0k}), at $n \to 0$ we obtain the
following in-medium renormalization for $\varphi^{(0)}(r)$ at short
distances:
\begin{equation}
\varphi(r)\simeq
   \varphi^{(0)}(r)\Bigl(1+\frac{8\sqrt{na^3}}{\sqrt{\pi}}\Bigr).
\label{varphi}
\end{equation}
Equation~(\ref{varphi}) is indeed a short-range approximation, and this
can be understood from the fact that, at sufficiently large momenta
$k$, the main contribution in the integral in the right-hand side of
Eq.~(\ref{LSch}) comes from the large momenta $q$. As $\lim_{q \to
\infty}U(q)/\widetilde{T}_{q}=0$, then for $k\to\infty$ Eq.~(\ref{LSch})
is also (like at $n \to 0$) reduced to the two-body Lippmann-Schwinger
equation, and, thus, $\varphi(r)$ obeys the Schr\"odinger equation
(\ref{twobody}) at small boson separations.  In order to explain the origin
of the factor $C=1+8\sqrt{na^{3}}/ \sqrt{\pi}$ in Eq.~(\ref{varphi}), we
remind that the boundary conditions (\ref{bc}), involved implicitly in the
in-medium Lippmann-Schwinger equation (\ref{LSch}), are valid for $r \gg
r_{0}$, where $r_{0}=n^{-1/3}$ is the mean distance between two particles in
the gas. The wave function $\varphi(r)$, related to a couple of bosons in the
Bose-Einstein condensate, does obey Eq.~(\ref{twobody}) at $r \to 0$, but it
differs from $\varphi^{(0)}(r)$, subjected to the boundary condition
(\ref{bound}), by the factor that is determined by the in-medium effects.
Thus, the range of validity of Eq.~(\ref{varphi}) is restricted by the region
$r\lesssim r_{0}$; in the vicinity of $r_{0}$ the behaviour of $\varphi(r)$
is rather complicated; and for $r\gg r_{0}$ the function $\varphi(r)$ is
subjected to the following asymptotics:
\begin{equation}
\psi(r)=\varphi(r)-1\simeq -\frac{1}{2{\pi}^{3/2}}\sqrt{\frac{a}{n}}
                                                 \frac{1}{r^{2}}.
\label{asymp}
\end{equation}
The latter evaluation can be obtained from Eq.~(\ref{psi}), which yields
$\psi(k) \simeq -\sqrt{\pi a/n}/k$ for $k\to 0$, and at $r\to\infty$ we
arrive at Eq.~(\ref{asymp}), contrary to the two-body problem, which implies
$\psi^{(0)}(r)\simeq -a/r$. Thus, the unusual ``overscreening" takes place
for the wave function $\varphi(r)$ owing to in-medium effects.

Now we are able to calculate the pair distribution function for a
dilute Bose gas from Eq.~(\ref{gpw}). By means of the substitution
(\ref{subst}) in the integral in Eq.~(\ref{gpw}) one can rewrite
the pair distribution function for $n\to 0$ in the form
\begin{equation}
g(r)\simeq\left(1+2\frac{n-n_0}{n}\right)\varphi^{2}(r).
\label{grnto0}
\end{equation}
where the relation (\ref{phiphi0}) is implied. This is short-range
approximation for $g(r)$, for the oscillating function $\cos({\bf k}{\bf
r})$ makes an important contribution to the integral in Eq.~(\ref{gpw})
at large $r$. To obtain a more concrete information from Eq.~(\ref{grnto0}),
one should calculate the condensate depletion.  It can be derived from
Eq.~(\ref{md}) with the ``scaling" substitution given by Eq.~(\ref{subst}).
This leads to
\begin{equation}
\frac{n-n_0}{n}=\int \frac{\d^{3}q}{(2\pi)^{3}}\frac{n_{q}}{n}
=\frac{8\sqrt{na^3}}{3\sqrt{\pi}}+\cdots
\label{depletion}
\end{equation}
The result again coincides with that of the effective-interaction
scheme because the momentum distribution (\ref{md}) is very close to
$n_k$ found within the modified Bogoliubov model. Rewriting
Eq.~(\ref{grnto0}) with the help of Eqs.~(\ref{varphi}) and
(\ref{depletion}), we arrive at the expression (\ref{grnto01}) that
has earlier been obtained by means of the theorems (\ref{HF}) and
(\ref{delU0}). One can see that Eq.~(\ref{grnto01}) is also violated
at the mean distance between particles, $r_{0}$, since
Eq.~(\ref{varphi}) for $\varphi(r)$ is broken when $r\gtrsim r_{0}$.

For strongly singular potentials, when $\varphi^{(0)}(r=0)=0$, the
correct result $g(r=0)=0$ takes place according to
Eq.~(\ref{grnto01}). As one can see from (\ref{grnto01}), the
correlation hole coming from the repulsion of bosons at small
particle separations gets less marked with an increase of the density
of the surrounding bosons, which is consistent with usual
expectations concerning the particle spatial separations.

Now it is interesting to compare the result given by
Eq.~(\ref{grnto01}) with the data of the Monte-Carlo calculations for
the hard-spheres~\cite{Giorg}. With the two-body wave function
(\ref{hardsphi0}) for the hard-sphere interaction (\ref{hards}),
Eq.~(\ref{grnto01}) reads
\begin{equation}
g(r)=
\left\{\begin{array}{ll}
0,                                            & r\leq a \\
(1-a/r)^{2}[1+64\sqrt{na^3}/(3\sqrt{\pi})],   & r>a.
\end{array}\right.
\label{grhards}
\end{equation}
\begin{figure}[tb]
\epsfxsize=85mm \centerline{\epsfbox{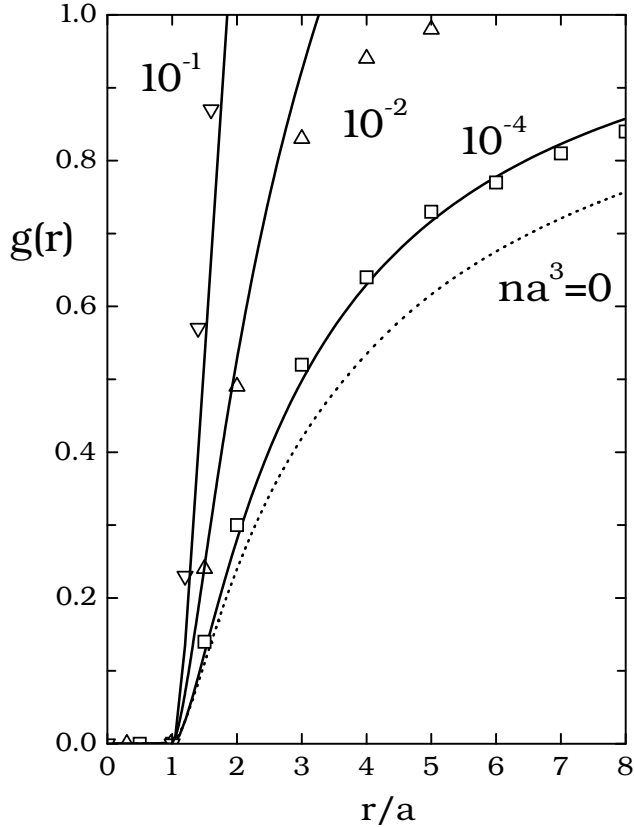}}

\caption{The pair distribution function corresponding to the hard-sphere
potential (\ref{hards}) for various values of the gas parameter $na^3$.
Solid lines -- the data of Eq.~(\ref{grhards}); $\Box$,
$\bigtriangleup$, $\bigtriangledown$ -- the data of the Monte-Carlo
calculations~\protect\cite{Giorg} for $na^3=10^{-4},\,10^{-2},\,10^{-1}$,
respectively.  The difference becomes significant at the mean distance
between particles $r_0$, where
$r_0/a=\left(na^3\right)^{-1/3}=21.5,\,4.64,\,2.15$, respectively.  Dotted
line -- Eq.~(\ref{grhards}) for $na^3=0$, one can see that $g(r)$ is
essentially changed even for small values of the gas parameter.}

\label{fig2}
\end{figure}
The results are displayed in Fig.~\ref{fig2} for various values of the gas
parameter $na^3$.  We remark that, strictly speaking, only the situation when
$na^3=10^{-4}$ can be investigated with Eq.~(\ref{grhards}) because even at
$na^3=10^{-2}$ we have $64\sqrt{na^3}/(3\sqrt{\pi})\approx 1.20$, so that the
next-to-leading order in Eq.~(\ref{grnto01}) makes contribution more
essential than that of the leading order in $na^3$. However, as it is seen,
the short-range approximation of Eq.~(\ref{grhards}) works well at small
values of $r$ even at $na^3=10^{-1}$. This agrees with a conclusion of
Giorgini et al.~\cite{Giorg} that Eq.~(\ref{eLY}), derived to the next-to-leading order in $na^3$, provides results
very close to the energy of a dilute Bose gas even at $na^3=10^{-1}$. Agreement
of the data of the Monte-Carlo calculations with the results coming from
Eq.~(\ref{gpw}) can be considered as {\it the direct positive test supporting
the expression (\ref{gpw}), which is based on the Bogoliubov principle of
correlation weakening in the strongly interacting Bose gas} (see Ref.~\cite{Ch1}).

Concluding this section, let us stress one more that
Eqs.~(\ref{varphi}) and (\ref{grnto01}) are correct only for
short boson separations. To determine the long- and intermediate-range behaviour of
$g(r)$ with the aim, for example, of finding the structural
factor $S(k)$ at all values of $k$, one should obtain $\varphi_{{\bf p}}({\bf r})$
for all values of ${\bf p}$ [remember that only the limiting
relation~(\ref{phiphi0}) have been utilized]. This
investigation is directly connected with the problem of the
relation between the boson momentum distribution and pair
scattering waves [see Eq.~(\ref{mdscw})] and needs additional
extended considerations being beyond the scope of this paper. The
same concerns the spectrum of elementary excitations. Now, and
here the point is, this problem is much more complicated as
compared to the Bogoliubov model (original or modified).
Within the Bogoliubov approach the
long-range behaviour of $g(r)$ is only governed by the two
quantities $\psi(k)=\langle a_{{\bf k}}a_{-{\bf k}}\rangle/n_{0}$ and
$n_k=\langle a^{\dagger}_{\bf k}a_{\bf k}\rangle$ [see Eq.~(\ref{gpwB})].
But now we have the rather complicated functional (\ref{gpw}) of $n_{k}$,
$\psi(k)$ and $\psi_{{\bf p}}({\bf k})$ given by Eq.~(\ref{25}).

\section{Thermodynamics of a dilute Bose gas}
\label{lowden}

After the necessary preparations, we are able to turn to the
thermodynamics of a strong-coupling Bose gas. The most simple way
of doing so is to deal with the chemical potential $\mu$ starting
from the well-known relation
\begin{equation}
\mu=\frac{1}{\sqrt{n_0}}\int \d^3r'\,\Phi(|{\bf r}-{\bf r}'|)
\langle \psi^{\dagger}({\bf r}')\psi({\bf r}')\psi({\bf r})\rangle,
\label{mu}
\end{equation}
valid in the presence of the Bose condensate~\cite{Bog2}. Here
$\psi^{\dagger}({\bf r})$ and $\psi({\bf r})$ stand for the Bose
field operators. This relation follows from the expression for
the infinitesimal change of the grand canonical potential
$
\delta\Omega=\langle\delta(\hat H-\mu\hat N)\rangle
$
and the necessary condition of the minimum for $\Omega$ with respect
to the order parameter $N_{0}$: $\partial\Omega(N_0,\mu,T)/\partial
N_0=0$, the Hamiltonian depending on the number of the condensed
particles owing to the substitution $a_0^{\dagger}=a_0= \sqrt{N_0}$.
Equations~(\ref{25}) and (\ref{mu}) lead to
\begin{equation}
\mu=n_0 {U}(0)
+\sqrt{2}\int\frac{\d^3q}{(2\pi)^3}n_q{U}_{{\bf q}/2}({\bf q}/2),
\label{mu1}
\end{equation}
where the quantity
\begin{equation}
{U}_{{\bf p}}({\bf k})=\int\d^3r\,\varphi_{\bf p}({\bf r})\Phi(r)
\exp(-i{\bf k}{\bf r})
\label{Up}
\end{equation}
is introduced. The quantities $U(k)$ and ${U}_{{\bf p}}({\bf k})$,
which can be called the in-medium scattering amplitudes, differ from
$f({\bf k},{\bf k}')$ introduced by Beliaev. The latter is determined
by the solutions of the ordinary two-body problem with the boundary
conditions corresponding to the usual plane-waves for $r \to \infty$. As
to $U(k)$, it is determined from the in-medium Lippmann-Schwinger
equation (\ref{LSch}).  In order to obtain the next-to-leading order
for $\mu$, it is sufficient to use the relation $\sqrt{2}
U(k)=\lim_{p \to 0}{U}_{{\bf p}}({\bf k})$, which follows from
Eq.~(\ref{phiphi0}). In this way, employing the substitution
(\ref{subst}) in the integral and taking into consideration
Eqs.~(\ref{U0}), (\ref{UkU0k}) and (\ref{depletion}), we can rewrite
Eq.~(\ref{mu1}) for $n \to 0$ in the form
\begin{eqnarray}
\mu&=&nU(0)\left(1+\frac{n-n_0}{n}+\cdots\right)
\nonumber \\
&=&\frac{4\pi\hbar^2
an}{m}\left(1+\frac{32}{3\sqrt{\pi}}\sqrt{na^3}+\cdots\right).
\label{mu2}
\end{eqnarray}
Equation~(\ref{mu2}), together with the thermodynamic relation
$\mu=\partial (n\varepsilon(n))/\partial n$, yields the Yang-Lee's result
(\ref{eLY}).

It is worth noting that we can also use the direct way of
calculating $\varepsilon$ based on Eq.~(\ref{emean}).
However, one could come to a wrong conclusion if Eq.~(\ref{emean})
employed in conjunction with Eq.~(\ref{mdscwB}). To use this way, one
should go beyond Eq.~(\ref{mdscwB}) and take into account the density
correction to that relation,
\begin{equation}
n_{k}(n_{k}+1)=\left(1+2\frac{n-n_0}{n}\right)\psi^{2}(k),
\label{newcon}
\end{equation}
following from Eq.~(\ref{mdscw}) at $n \to 0$. Thus, the
preliminary result found in Ref.~\cite{Ch3} should be abandoned in
favour of Eq.~(\ref{eLY}).

The correct behaviour of the pair distribution function found in the
previous section allows for deriving the correct results for the
kinetic and interaction energies by a direct calculation, even beyond
the Hellmann-Feynman theorem. For example, substituting
Eq.~(\ref{grnto01}) into Eq.~(\ref{eint1}) and taking account of
Eq.~(\ref{da}), one can immediately find Eq.~(\ref{eint}).

\section{Conclusion}
\label{concl}

In conclusion, we remark that the present paper deals with the
thermodynamics of a Bose gas of strongly interacting particles in the
leading and next-to-leading orders of the expansion in $na^3$. The
strong-coupling generalization of the Bogoliubov model considered
here reproduces the well-known formula of Lee and Yang (\ref{eLY})
and, contrary to the effective-interaction approach of
Ref.~\cite{Lee}, yields correct results for the kinetic and interaction
energies and
short-range spatial correlations of bosons. To go further, additional
investigations should be fulfilled. In particular, it is necessary to
solve the problem concerning the relation that connects the boson
momentum distribution with the scattering waves. The spectrum of the
elementary excitations should also be considered within the approach
of Refs.~\cite{Ch,Ch1,Ch3} in order clarify to what extent it differs from
the well-known prediction of the effective-interaction approach. We,
of course, mean the region of intermediate momenta rather than the
linear phonon part, which should be the same according to the
thermodynamic prescription. This question can not be answered without
addressing the long-range spatial boson correlations.

The authors are grateful to S.~Giorgini for making the data of the
Monte-Carlo calculations~\cite{Giorg} available to us.
This work was supported by the RFBR grant No. 00-02-17181.



\begin{references}
\bibitem{classif} We employ the following classification of the short-range
pairwise potentials. The potential is of weak-coupling type if it is
integrable and proportional to a small parameter. In this case the Born
approximation works well in the ordinary two-body problem, and, hence, in the
Born series (\ref{expa}) $|a_{0}|\ll |a_{1}|$. Otherwise the potential is
called singular.  Finally, we deals with the strongly singular potential
(strong coupling regime) if the individual terms in Eq.~(\ref{expa}) do not
exist due to strong repulsion at short distances. For example, the potential
$\Phi(r)\propto 1/r^{m}$ $(m>3)$ at $r\to 0$ or the hard-sphere potential
(\ref{hards}) belongs to that type.
\bibitem{NoteB} Here the ordinary scattering problem for particles
with the mass $m/2$ ($m$ denotes the mass of the boson)
is implied so that $\varphi^{(0)}_{{\bf p}}({\bf r})$
tends to $\exp(i{\bf p}{\bf r})$ as $r \to \infty$. Whereas the
pair wave functions $\varphi_{{\bf p}}({\bf r})$ used in
Secs.~\ref{pairwaves}-\ref{lowden} are symmetrized with
respect to the boson permutation and, thus, obey the boundary
conditions~(\ref{scw}) and (\ref{bc}). So, the
scattering amplitude $U_{{\bf p}}({\bf k})$ used below is not the
same as $f({\bf k},{\bf p})$ and this is by two reasons. First, it is
due to symmetrization of the pair wave functions. Second,
$U_{{\bf p}}({\bf k})$ is governed by the {\it in-medium} pair wave
functions.
\end{references}
\end{document}